\documentclass[prl,twocolumn,notitlepage,longbibliography]{revtex4-1}
\usepackage{amsmath}
\usepackage{amssymb}

\usepackage[unicode=true,colorlinks=true,citecolor=blue,urlcolor=blue]{hyperref}

\usepackage{bm}
\usepackage{epsfig}

\usepackage[normalem]{ulem}

\renewcommand {\Im}{\mathop\mathrm{Im}\nolimits}
\renewcommand {\Re}{\mathop\mathrm{Re}\nolimits}

\renewcommand {\phi}{{\varphi}}
\newcommand {\rmi}{{\rm i}}
\newcommand {\rmd}{{\rm d}}

\newcommand {\e}{{\rm e}}
\newcommand {\eps}{\varepsilon}

\begin{document}
\title{
{Quantum Borrmann effect for  dissipation-immune photon-photon correlations}\\
}

\author{Alexander V. Poshakinskiy}
\affiliation{Ioffe Institute, St. Petersburg 194021, Russia}

\author{Alexander N. Poddubny}
\email{poddubny@coherent.ioffe.ru}

\affiliation{Ioffe Institute, St. Petersburg 194021, Russia}

\begin{abstract}
We study theoretically the second-order correlation function $g^{(2)}(t)$ for  photons transmitted through a periodic Bragg-spaced array of superconducting qubits, coupled to a waveguide. We demonstrate that  photon bunching and anti-bunching persist much longer than both radiative and non-radiative lifetimes of a single qubit.  The photon-photon correlations become immune to non-radiative dissipation due to the  Borrmann effect, that is a strongly non-Markovian collective feature of light-qubit coupling inherent to  the Bragg regime.
This persistence of quantum correlations opens new avenues for enhancing the performance of   setups of waveguide quantum electrodynamics.
\end{abstract}
\date{\today}

\maketitle
{\it Introduction}. Cooperative  effects are widely used to manipulate light-matter interactions~\cite{KimbleRMP2018,Limonov2017}.  
Namely, constructive or destructive interference between light coupled to different  resonant emitters can result in enhancement  (superradiance) or suppression (subradiance) of the radiative decay rate  $\Gamma_{\rm rad}$  as compared to that of  an individual emitter. Both super- and sub-radiant modes have been demonstrated for a variety of experimental platforms, such as resonant  plasmonic~\cite{taubert2012} and  dielectric nanostructures~\cite{Kuznetsov2016}, solid-state quantum emitters~\cite{Averkiev2009,khitrova2011}, individual molecules, etc. For example, long-living supercavity modes with $\Gamma_{\rm rad}\ll \Gamma_0$, inspired by the photonic bound states in continuum~\cite{Hsu2016},  have been recently realized  for  resonant dielectric nanoparticles~\cite{Koshelev2020}. Similar concepts to engineer subradiant modes apply in the quantum regime~\cite{Ostermann_2019,Molmer2019,Ke2019,Poddubny2019quasiflat} as has been recently demonstrated 
for single-photon excitations of a superconducting qubit array coupled to a waveguide~\cite{brehm2020waveguide}. It is however much harder to suppress the {\it nonradiative} decay. In the usually  valid Markovian  regime of light-matter coupling, the nonradiative decay  rate $\Gamma_{\rm nonrad}$ just adds an independent contribution to the total decay rate, $\Gamma_{\rm tot}=\Gamma_{\rm rad}+\Gamma_{\rm nonrad}$ that is  not sensitive neither to the interference nor to  the number of emitters.  As such, the maximum lifetime  $1/(2\Gamma_{\rm tot})$ is given by $t^{(1)}_{\rm nonrad}=1/(2\Gamma_{\rm nonrad})$ which seems to limit the performance of a real-life quantum system regardless of the sophisticated techniques used to manipulate $\Gamma_{\rm rad}$.

\begin{figure}[b!]
\includegraphics[width=0.45\textwidth]{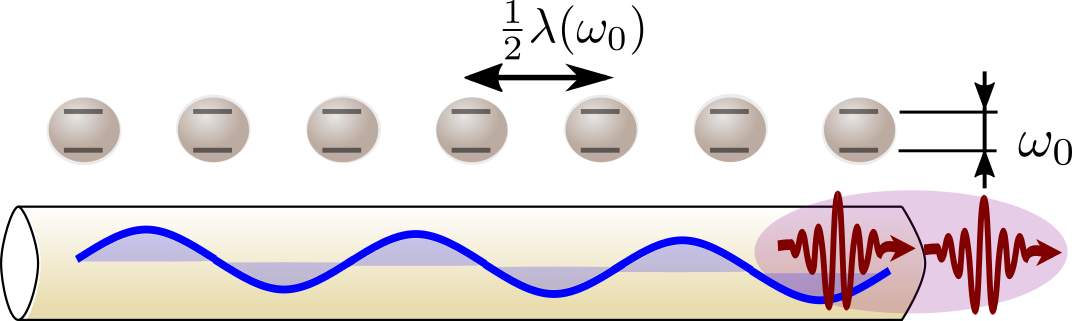}
\caption{Schematics of two photons propagating in an array of superconducting qubits coupled to a waveguide and separated by the Bragg spacing $d=d_{\rm Bragg}\equiv \lambda(\omega_0)/2$. 
   }\label{fig:1}
\end{figure}

In this Letter we propose a simple scheme  to achieve quantum  correlations between photons in an array of 
 superconducting qubits in a waveguide that have  the total lifetime much larger that {\it both} radiative  and nonradiative lifetime of an individual qubit.
 We consider a periodic array of $N$ two-level qubits with the resonant frequency $\omega_0$ and the spacing $d$, satisfying the resonant Bragg condition
 \begin{equation}
d=d_{\rm Bragg}\equiv \lambda(\omega_0)\frac{m}{2},\quad m=1,2,\ldots\:,
 \end{equation}
where $\lambda(\omega_0)=2\pi c/\omega_0$ and $c$ is the speed of light, see Fig.~\ref{fig:1}.
 Our proposal is inspired by the experimentally discovered in 1950 but not very widely known Borrmann effect~\cite{borrmann1950}. The Borrmann effect manifests itself in anomalous transmission of X-rays through crystals due to suppression of their absorption  in the Bragg regime, when the electromagnetic wave has nodes at the atom centers. Similar ideas were examined in the  classical optics of  Bragg arrays of semiconductor quantum wells~\cite{Ivchenko1994,koch1996mqw,Goldberg2009}, in particular Refs.~\cite{Poshakinskiy2012,Ivchenko2013}. However,  to the best of our knowledge  the Borrmann effect has never been studied in the quantum regime, that is relevant for recently emerging  setups of waveguide quantum electrodynamics based on  cold atoms and superconducting qubits~\cite{Corzo2016,Corzo2019,Mirhosseini2019}.

Here we perform a rigorous calculation of the  second-order photon-photon correlation function $g^{(2)}(t)$ and demonstrate long-living  bunching [$g^{(2)}(t)>1$] and antibunching  [$g^{(2)}(t)<1$] of photons transmitted through the qubit arrays at the times $t\gg t^{(1)}_{\rm nonrad}$. 
The correlations that persist much longer than the {\it radiative} lifetime are already known for a  two-qubit system separated by large anti-Bragg distance $d=(m\pm \frac{1}{2})\lambda_0/2$~\cite{Baranger2013}. However, the advantage of current proposal based on the multi-qubit Bragg array is that the correlations survive at even longer times, exceeding the {\it non-radiative} lifetime of a single qubit. This could open new possibilities for applications in quantum memory and quantum information processing.


\begin{figure}[t!]
\includegraphics[width=0.45\textwidth]{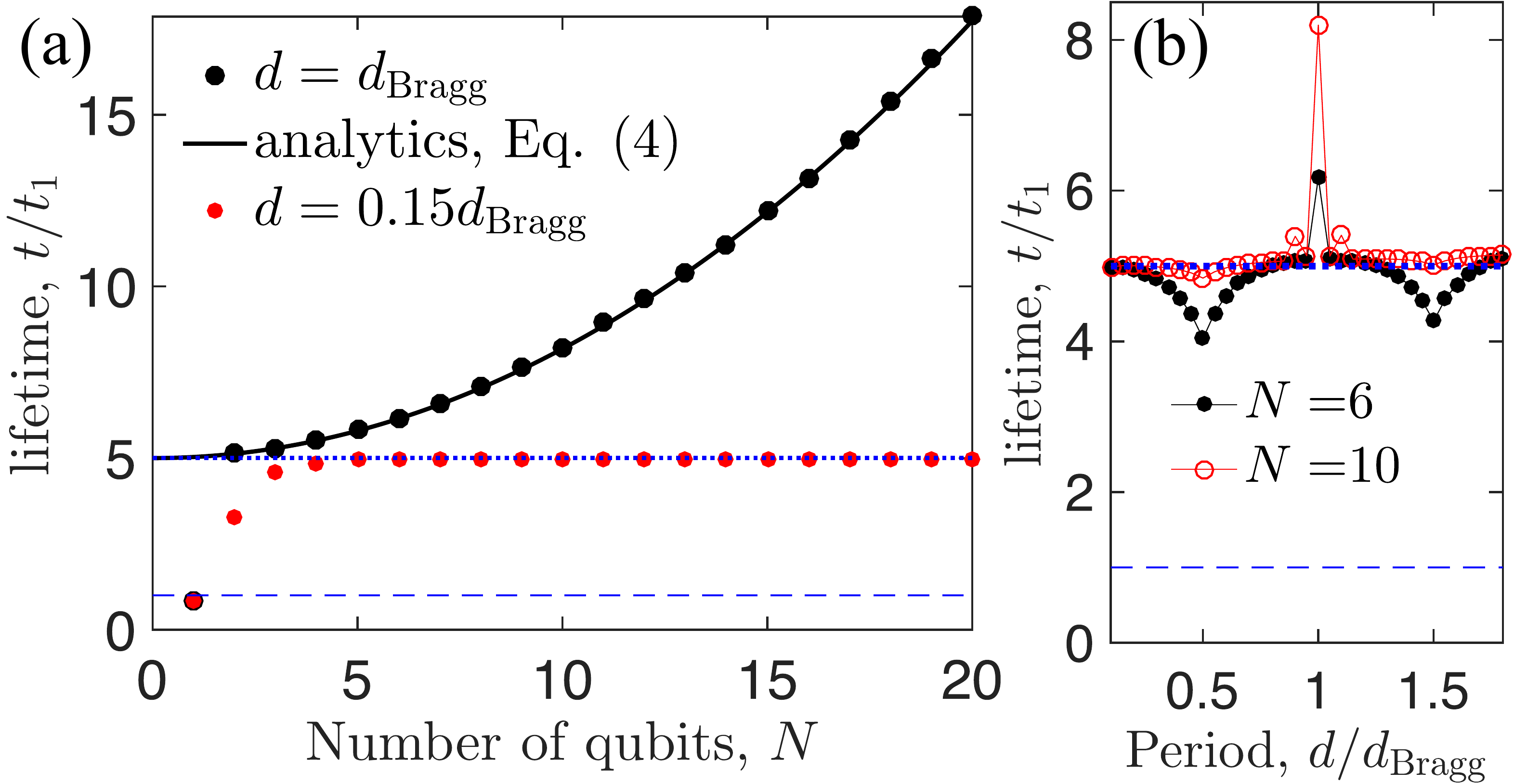}
\caption{  (a) Dependence of the lifetime of longest-living eigenmode on the number of qubits for a Bragg-spaced array with $\omega_0d/c=\pi$ (large black symbols)
and a short-period metamaterial  with $\omega_0d/c=0.15\pi$ (small red symbols).  Solid curve shows the analytical dependence Eq.~\eqref{eq:time}.
(b)  Lifetime dependence on the array period. Black filled and red open circles correspond to $N=6$ and $N=10$ qubits.
Blue dotted and dashed lines show the nonradiative and radiative lifetimes for a single qubit, respectively.
Calculation has been performed for $\Gamma_{\rm nonrad}={0.2}\Gamma_{0}$  and $\Gamma_0/\omega_0=10^{-2}$.   
   }\label{fig:2}
\end{figure}

{\it Complex eigenfrequencies.}
We will now analyze  the energy spectrum  of single-excited states of photons coupled to the qubits in a waveguide that will determine the lifetime of photon-photon correlations. 
The values of complex eigenfrequencies $\omega$ are determined by  the positions of the poles of the matrix Green function 
\begin{equation}\label{eq:G}
G(\omega)=[H(\omega)-\omega]^{-1}
\end{equation}
that is defined from the Hamiltonian matrix~\cite{Ivchenko2005,Baranger2013,Fang2014,Ke2019}
\begin{equation}\label{eq:H}
H_{mn}(\omega)=(\omega_0-\rmi \Gamma)\delta_{mn}-\rmi \Gamma_0\e^{\rmi \omega |z_m-z_n|/c},\quad m,n=1\ldots N\:,
\end{equation}
where $z_m$ are the coordinates of the qubits. Crucially, we do not limit ourselves to the Markovian approximation and take into account that the matrix $H_{mn}$ does depend on the frequency $\omega$ via the phase $\omega |m-n|d/c$ gained by light when traveling from the qubit $m$ to the qubit $n$. As a result of the breaking down of Markovian approximation, the eigenfrequency equation  $\det G^{-1}(\omega)=0$ for the  $N$-qubit system acquires an infinite number of eigenvalues. In addition to the $N$ eigenvalues obtained in the Markovian approximation $H(\omega)=H(\omega_0)$, there exists also an infinite number of Fabry-Per\'ot eigenmodes that approach $\omega_0$ with the increase of $N$.

We start by analyzing the lifetime of the longest living eigenmode $t=1/(2\min |\Im \omega|)$. The lifetime dependence on the number of qubits $N$   for the Bragg-spaced array is shown by the black circles in Fig.~\ref{fig:2}(a). The lifetime greatly exceeds the nonradiative lifetime of a single qubit (blue dotted line).  
The lifetime dependence  is well described by the analytical equation
\begin{equation}\label{eq:time}
\frac{t(N)}{t_{\rm nonrad}^{(1)}}=1+\frac{2\Gamma_0}{\omega_0 \pi}N^2\:,
\end{equation}
shown by the solid black curve in  Fig.~\ref{fig:2}(a).   This quadratic  $t(N)$ dependence is very different from the case of short-period quantum metamaterial  with the spacing defined by $ \omega_0d/c=0.15\pi$, where the Markovian approximation works well and  the lifetime is limited from above by $1/\Gamma_{\rm nonrad}$, small red circles in Fig.~\ref{fig:2}(a).  The increase of the lifetime in Bragg case is also seen in Fig.~\ref{fig:2}(b) where we plot its dependence on the array period for two given numbers of qubits $N=6$ and $N=10$. For $N=6$ the Markovian approximation still works and the lifetime weakly depends on period. However, already for $N=10$ qubits the lifetime in Bragg structure  increases,  evidencing strongly non-Markovian physics.

\begin{figure}[t!]
\includegraphics[width=0.45\textwidth]{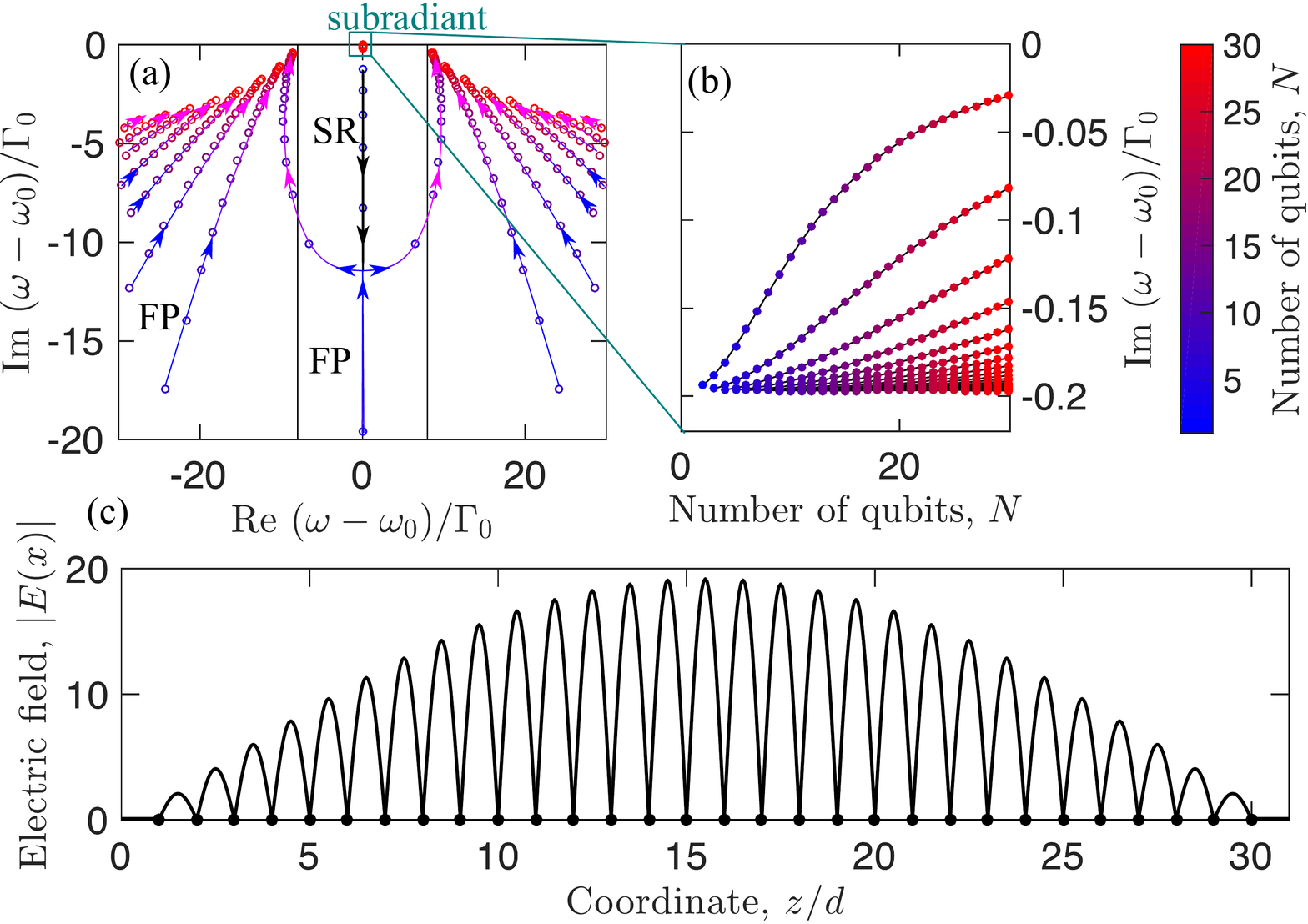}
\caption{
(a) Evolution of the complex energy spectrum of the Bragg qubit array with the number of qubits $N$. Lines with arrows show the  the trajectories of 
Fabry-Per\'ot (FP) and superradiant (SR)  modes with the increase of $N$. 
(b) Blow-up of the subradiant part of the  spectrum depending on $N$. (c) Electric field Eq.~\eqref{eq:E} calculated for the most subradiant mode of the 30-qubit array ($\omega-\omega_0\approx -0.015\Gamma_0$).
Calculation has been performed for $\Gamma_{\rm nonrad}=0.2\Gamma_{0}$ and $\Gamma_0/\omega_0=10^{-2}$.  
}\label{fig:3}
\end{figure}

Details of the evolution of the complex energy spectrum of the Bragg array  with the number of qubits $N$ are examined in Fig.~\ref{fig:3}. In the Markovian approximation the energy spectrum includes a superradiant mode with $\omega_{\rm SR}=\omega_{0}-\rmi (N\Gamma_0+\Gamma)$ and $N-1$ degenerate dark modes,
$\omega_{\rm dark}=\omega_{0}-\rmi \Gamma$. Figure~\ref{fig:3}(a) shows the evolution of spectrum with increasing $N$.  The trajectory of the superradiant mode in the complex plane  is shown by the vertical black line in Fig.~\ref{fig:3}(a). At $N=6\sim\sqrt{\omega_0/\Gamma_0}$ the structure exhibits a transition from the superradiant regime to the photonic crystal regime. Namely, the superradiant mode collides with the another  mode with $\Re \omega=\omega_0$ ( vertical  blue line in Fig.~\ref{fig:3}a). After the collision  these two modes split into a pair of Fabry-Per\'ot modes, that are mirror-symmetric with respect to $\omega_0$. 
For larger $N$, the spectrum consists of multiple Fabry-Per\'ot modes. Their trajectories are shown by  {curved colored lines  with arrows} in Fig.~\ref{fig:3}(a).
In the limit $N \to \infty$, the Fabry-Per\'ot modes condense near the two branching points of the Green's function of the infinite qubit array
\begin{equation}\label{eq:gap}
\omega_{\pm}=\omega_0\pm \sqrt{\Delta^2-\frac{\Gamma^2}4} - \frac{\rmi\Gamma}2,
\end{equation}
where, $\Delta=\sqrt{2\Gamma_0\omega_0/\pi}$ is the halfwidth the polariton band gap~\cite{Deych2000,SpecFreqs2}.  Note that the decay rates of the Fabry-Per\'ot modes remain finite and tend to $-\Im\omega_\pm\equiv \Gamma/2$, a half of the nonradiative decay rate of the qubits, reflecting the  half-light half-qubit excitation nature of the polariton.  
The specific feature of the Bragg array is that the band gap width $2\Delta$ exceeds the radiative linewidth of a single qubit  $\Gamma_0$ by a large factor $\sim\sqrt{\omega_0/\Gamma_0}$.

 In present work, we are more interested in the evolution of the subradiant modes in the vicinity of the resonance $\omega_0$. The real parts of their eigenfrequencies  are equal to $\omega_0$ and the evolution of the imaginary parts with $N$ is shown in Fig.~\ref{fig:3}(b). The calculation demonstrates that when the number of qubits increases, the spectrum of subradiant modes is no longer degenerate. Moreover, their imaginary part becomes less than $\Gamma_{\rm nonrad}$ by the absolute value which means that  the effective lifetime of the eigenstates $t=1/(2|\Im \omega|)$ becomes longer than both nonradiative and radiative lifetimes of a single qubit. 
In the limit $N \to \infty$, the subradiant modes condense near  $\omega = \omega_0 - \rmi\Gamma$ and $\omega = \omega_0$. The latter means the presence of modes with infinitely long lifetime.  
 In order to  understand qualitatively the origin of the long lifetime we plot in Fig.~\ref{fig:3}(c) the distribution of the electric field, corresponding to the emission from the longest living eigenmode for $N=30$ qubits:
\begin{equation}\label{eq:E}
E(z)=-\rmi\sum\limits_{m=1}^N \e^{\rmi \omega |z-md|/c}\psi_m\:,
\end{equation}
where $\psi_m$ is the eigenvector satisfying  $H(\omega)\psi=\omega \psi$. The calculated distribution  has nodes at the qubit sites, which explains the suppression of the nonradiative decay, similar to the Borrmann effect in the X-ray physics~\cite{borrmann1950}.
In another words, the mixed light-qubit polariton wave decouples from the qubits and stops being absorbed.


{\it Persistent quantum correlations.}
The uncovered long-lived modes pave the way for the long-lived quantum correlations in the Bragg qubit array, with the decay times longer than the nonradiative decay rate. In order to demonstrate this, we calculate the photon-photon correlation function $g^{(2)}(\tau)=\langle a^\dag (0) a^\dag (\tau) a(\tau) a (0) \rangle/\langle a^\dag (0) a(0)\rangle $, where $a(t)$ is the photon destruction operator,  for the light transmitted through the Bragg array under low-intensity coherent excitation at frequency $\eps$.
The correlation function can be calculated as  \cite{Baranger2013,Ke2019}:
\begin{equation}\label{eq:g2t}
g^{(2)}(\tau,\varepsilon)=
\left| 1+\frac{\rmi}{2t^2(\varepsilon)}\int\limits_{-\infty}^\infty \frac{\rmd \omega}{2\pi}\e^{-\rmi \omega \tau}M(\varepsilon+\omega,\varepsilon-\omega)\right|^{2},
\end{equation}
where
$
t=1+\rmi \Gamma_0 \sum_{m,n=1}^{N}G_{mn}(\varepsilon)\e^{\rmi \eps(z_{m}+z_{n})/c}
$
is the single-photon transmission coefficient and the kernel
\begin{equation}\label{eq:M}
M(\omega_1',\omega_2')= -2\rmi \Gamma_0^2\sum_{m,n=1}^{N} s_n^-(\omega_1')s_n^-(\omega_2')Q_{nm}s_m^+(\eps)s_m^+(\eps) \end{equation}
with $s_m^\pm(\omega) = \sum_m G_{mn} \e^{\pm\rmi (\omega/c) z_n}$ and  $Q_{nm} =[\Sigma^{-1}]_{nm}$, 
 $\Sigma_{nm}=\int\rmd\omega G_{nm}(\omega)G_{nm}(2\eps-\omega)/(2\pi)$
describes photon-photon interaction in the qubits. The integration over $\omega$ in Eq.~\eqref{eq:g2t} in the non-Markovian regime requires some care due to the presence of factors $\e^{\pm\rmi (\omega/c) z_n}$ in $s^\pm$. In order to simplify these factors, we take into account the identity 
\begin{equation}
s_m^\pm(\omega)= -\frac{\e^{\rmi (\omega/c) z_{n^\pm}}}{\rmi\Gamma_0} \left[ \delta_{m,n^\pm} + (\omega-\omega_0+\rmi \Gamma) {G}_{m,n^\pm} \right]
\end{equation}
where $n^+=1$, $n^-=N$  that follows from the Green function definition Eqs.~\eqref{eq:G}--\eqref{eq:H}. Next, we expand the Green function as
%
$G_{mn}=\sum_\nu g^\nu_{mn}/(\omega_\nu-\omega)$
where $\omega_\nu$ are the eigenfrequencies of the Hamiltonian Eq.~\eqref{eq:H} that are found numerically within some large finite region and the residue matrices $g^\nu_{mn}$ are determined following Ref.~\cite{Gippius2009}.  Finally, using the sum rule 
$\sum_\nu g^\nu_{mn}=\delta_{mn}$, we find
\[
s_{m}^\pm=\sum\limits_{\nu}\frac{s^{\nu,\pm}_m}{\omega_\nu-\omega},\:\:\:
s_m^{\nu,\pm}=\operatorname{i}\e^{\rmi (\omega/c) z_{n^\pm}}g_{m,n^\pm}^\nu\frac{\omega_\nu-\omega_0+\rmi \Gamma}{ \Gamma_0} \:.
\]
After  this expansion is substituted into Eq.~\eqref{eq:M}, we use standard contour integration techniques to obtain the correlation function Eq.~\eqref{eq:g2t}.
\begin{figure}[t!]
\includegraphics[width=0.45\textwidth]{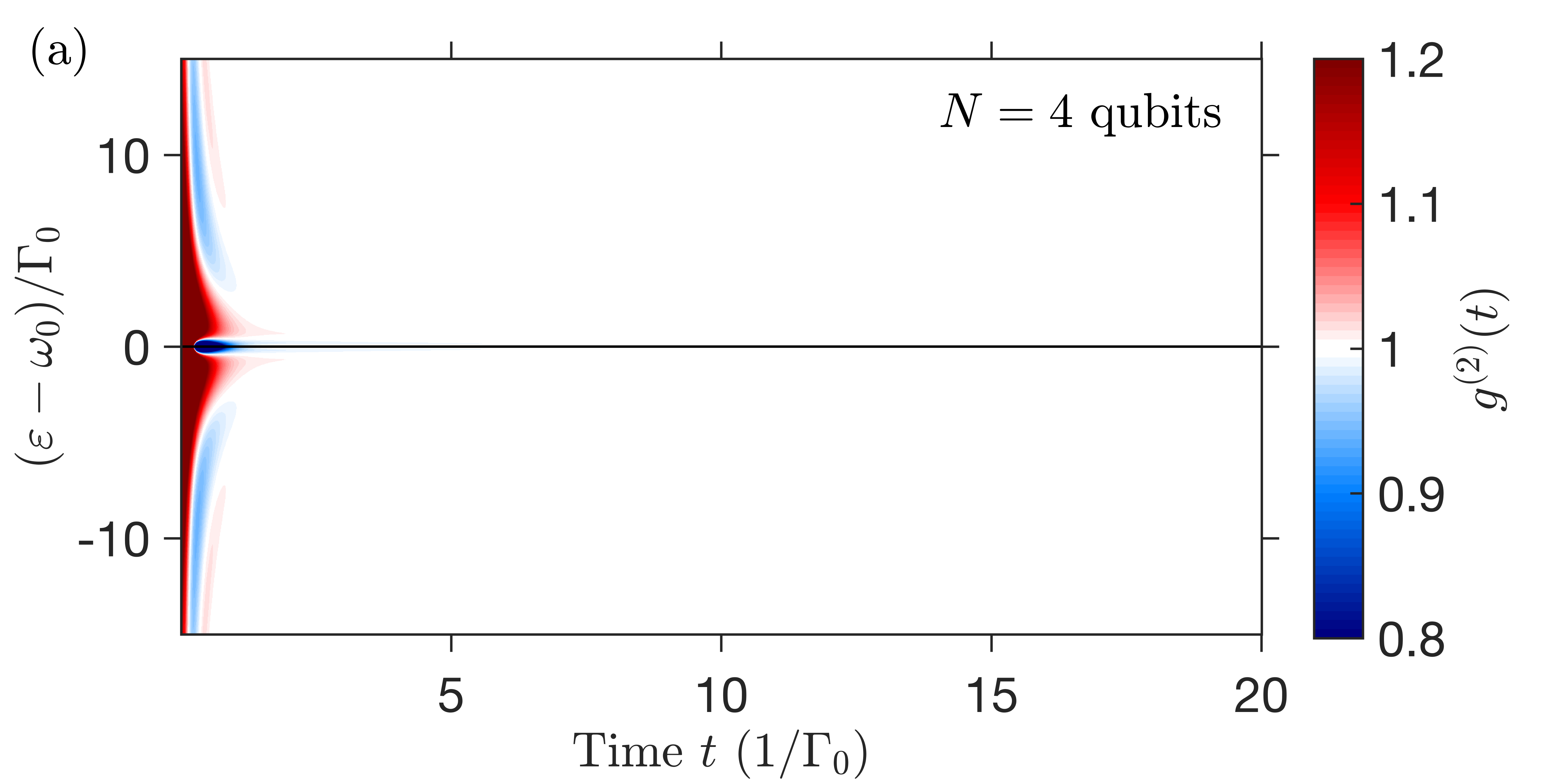}
\includegraphics[width=0.45\textwidth]{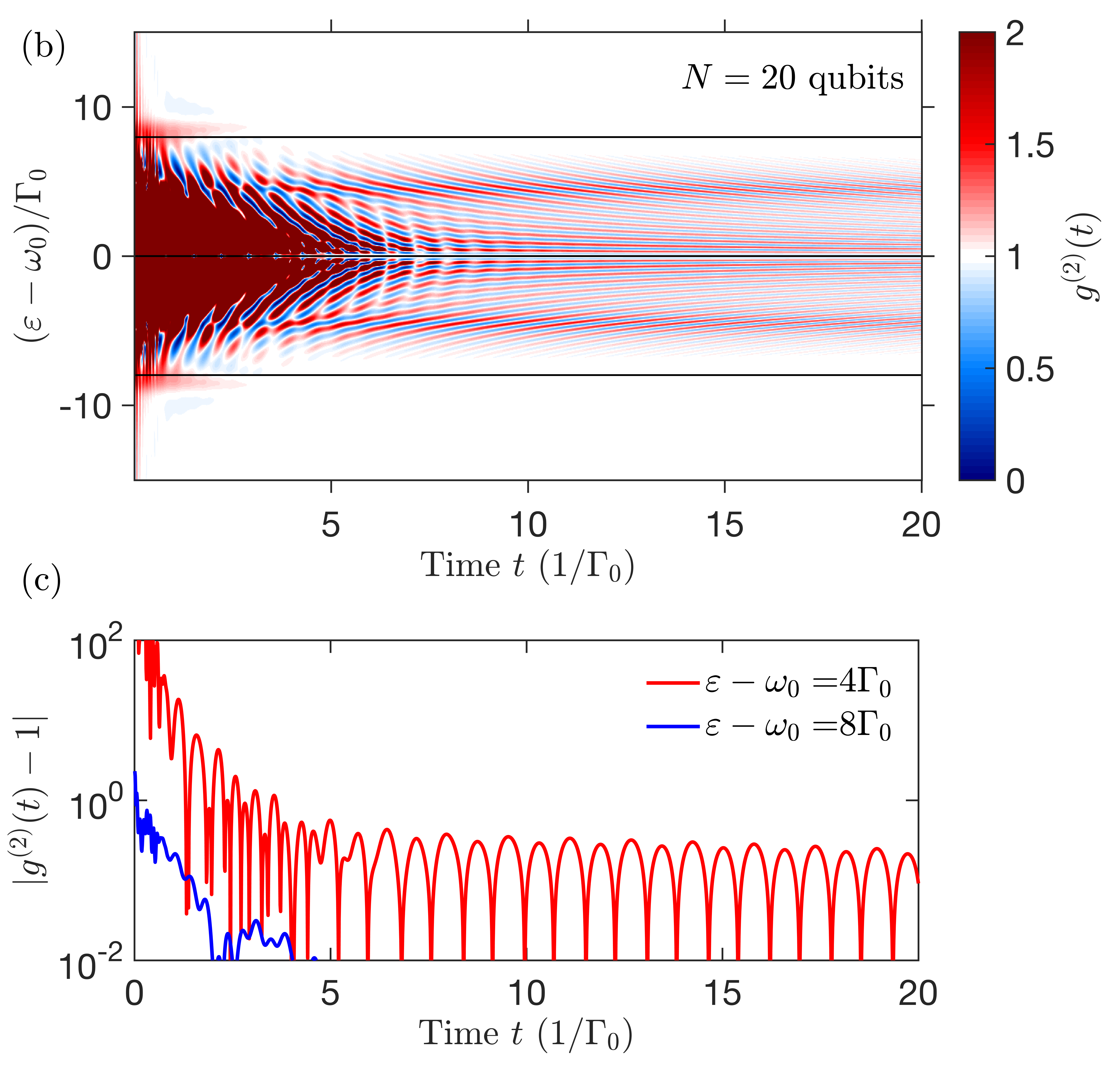}
\caption{(a,b) Time dependence  of photon-photon correlations $g^{(2)}(t)$ scanned vs. the energy of incoming photons $\eps$ calculated for the Bragg array with (a) $N=4$ qubits and (b) $N=20$ qubits. 
Upper and lower horizontal lines in (b) indicate the edges of the polariton bandgap $\eps=\omega_0\pm \Delta$. 
Panel (c) shows the time dynamics for $N=20$ qubits at two excitation characteristic energies. Calculation has been performed for $\Gamma=0.2\Gamma_0$. 
}\label{fig:4}
\end{figure}

Figure~\ref{fig:4} shows the time dependence $g^{(2)}(\tau)$ obtained numerically for different incident light frequencies $\eps$. 
Panels (a) and (b) have been calculated for a short array with $N=4$ qubits that is still in the superradiant regime and for a longer array with $N=16$ qubits that is already in the photonic crystal regime, respectively. At relatively short times the function $g^{(2)}(\tau)$ demonstrates strong photon bunching  when the excitation frequency $\eps$ is close to $\omega_0$. The single photon transmittance $|t(\omega_0)|^2$   is suppressed  in the vicinity of resonance due to the strong reflection,  so that photons can pass through the structure only in pairs. 
The calculation demonstrates that photon-photon correlations strongly depend on the number of qubits. In the case of short structure, Fig. \ref{fig:4}(a), in the wide spectral range $\Gamma_0 \lesssim |\varepsilon-\omega_0| \lesssim N\Gamma_0$, the function $g^{(2)}(\tau)$ rapidly decays to $1$ at the time scale of the superradiant mode $\sim 1/(N\Gamma_0)$. 
In long structures, Fig. \ref{fig:4}(b), the decay is non-monotonous and the correlations oscillate with time. 
Strong photon bunching is observed when the excitation energy is close to the edges of the polariton band gap
$\omega_0\pm\Delta\approx \omega_0\pm 8\Gamma_0$, shown by the horizontal lines in Fig. \ref{fig:4}b.
Crucially, the  decay of the correlations becomes significantly slower. This is also seen from  Fig.~\ref{fig:4}(c),  showing the dynamics of the correlations for two values of the   excitation energies $\eps=\omega_0+ 4\Gamma_0$ and $\eps=\omega_0+8\Gamma_0$  respectively. The red curve in  Fig.~\ref{fig:4}(c) is calculated for the excitation energy 
$\eps=\omega_0+ \Delta/2\approx \omega_0+4\Gamma_0$, when the amplitude of the correlations at large times is at maximum.
This   corresponds to the regime of photon-photon interaction  when one of the two scattered photons is at the resonance with the Fabry-Per\'ot mode near the polariton band gap edge $ \omega_0+\Delta$ and at the same time the other photon is at the resonance with the subradiant states with $\Re\omega= \omega_0$.
  The lifetime of the correlations is longer than both radiative and nonradiative lifetimes of a single qubit $t_{\rm rad}=1/(2\Gamma_0)$ and $t_{\rm nonrad}=5t_{\rm rad}$  for the parameters of Fig.~\ref{fig:4}. According to Fig.~\ref{fig:2}(a) and Eq.~\eqref{eq:time}, the lifetime grows for longer arrays as $N^2$. Since $\text{min}(\Im \omega)\to 0$ for $N\to \infty$, we expect slow nonexponential power law decay in the limit of infinite structure.

To summarize, we predict that photon-photon correlations become partially immune from nonradiative dissipation due to the Borrmann effect, when the light wave has nodes at the qubit positions. Our results demonstrate that 
the properties of the Bragg-spaced array of qubits, when $\omega_0d/c=\pi,2\pi,\ldots$ are strongly different from those in the conventional metamaterial regime when $\omega_0d/c\ll \pi$~\cite{brehm2020waveguide}. Thus, the Bragg-spaced arrays offer new possibilities to manipulate the quantum light.   We also note, that the time dynamics $g^{(2)}(t)$, studied here, probes directly only the lifetimes of single-excited states. Even more interesting physics can be expected for the double-excited states.
The interaction-induced localization and topological transitions have been recently predicted for two-polariton states~\cite{Zhong2020,alex2020quantum}, but the  non-Markovian regime of polariton-polariton interactions remains fully unexplored and promises very intriguing effects.
  For instance, one can imagine a situation when a pass-band of bound two-polariton states ~\cite{Zhang2019arXiv,Rabl2020} forms within a wide Bragg band gap of single-particle excitations, which could lead to the highly selective two-photon transmission.

\begin{acknowledgments}
We acknowledge numerous useful discussions with E.L.~Ivchenko and support from   the Russian President Grant No. MD-243.2020.2.
\end{acknowledgments}

%

\end{document}